\documentclass[a4paper]{article}

\usepackage{INTERSPEECH2020}
\usepackage{xcolor}
\usepackage[flushleft]{threeparttable} 

\title{Efficient neural speech synthesis for low-resource languages through multilingual modeling}
\name{Marcel de Korte, Jaebok Kim, Esther Klabbers}
\address{
  ReadSpeaker \\ Huis ter Heide, the Netherlands}
\email{\{marcel.korte,jaebok.kim,esther.judd\}@readspeaker.com}

\begin{document}

\maketitle

\begin{abstract}
Recent advances in neural TTS have led to models that can produce high-quality synthetic speech. However, these models typically require large amounts of training data, which can make it costly to produce a new voice with the desired quality. Although multi-speaker modeling can reduce the data requirements necessary for a new voice, this approach is usually not viable for many low-resource languages for which abundant multi-speaker data is not available. In this paper, we therefore investigated to what extent multilingual multi-speaker modeling can be an alternative to monolingual multi-speaker modeling, and explored how data from foreign languages may best be combined with low-resource language data. We found that multilingual modeling can increase the naturalness of low-resource language speech, showed that multilingual models can produce speech with a naturalness comparable to monolingual multi-speaker models, and saw that the target language naturalness was affected by the strategy used to add foreign language data.

\end{abstract}
\vspace{0.1cm}

\section{Introduction}

Over the past few years, developments in sequence-to-sequence (S2S) neural text-to-speech (TTS) research have led to synthetic speech that sounds almost indistinguishable from human speech (e.g.~\cite{shen2018natural,kons2019high,ren2019fastspeech}). However, large amounts of high-quality recordings are typically required from a professional voice talent to train models of such quality, which can make them prohibitively expensive to produce. To counter this issue, investigations into how S2S models can facilitate multi-speaker data has become a popular topic of research~\cite{jia2018transfer,gibiansky2017deep,deng2018modeling}. 
A study by~\cite{latorre2019effect}, for example, showed that multi-speaker models can perform as well or even better than single-speaker models when large amounts of target speaker data are not available, and that single-speaker models only perform better when substantial amounts of data are used. Their research also showed that the amount of data necessary for an additional speaker can be as little as 1250 or 2500 sentences without significantly reducing naturalness. With regards to parametric synthesis,~\cite{luong2019training} investigated the effect of several multi-speaker modeling strategies for class imbalanced data. Their research found that for limited amounts of speech, multi-speaker modeling and oversampling could improve speech naturalness compared to single speaker models, while undersampling was found to generally have a harmful effect. They also showed that ensemble methods can further improve naturalness, but this strategy comes with a considerable computational cost that is usually not feasible for S2S modeling.

Although the above research shows that multi-speaker modeling can be an effective strategy to reduce data requirements, it is not a suitable solution for many languages for which large quantities of high-quality multi-speaker data are not available.
Multilingual multi-speaker synthesis aims to address this issue by training a multilingual model on the data of multiple languages. Among the first to propose a neural approach to multilingual modeling was~\cite{li2016multi}. Instead of modeling languages separately, they modeled language variation through cluster adaptive training, where a mean tower as well as language basis towers were trained. They found that multilingual modeling did not harm naturalness for high-resource languages, while low-resource languages benefited from multilingual modeling. Another study by~\cite{gutkin2017uniform} scaled up the number of unseen low-resource languages to twelve, and similarly found that multilingual models tend to outperform single speaker models.

More recently, multilingual modeling was also adopted in S2S architectures~\cite{nachmani2019unsupervised,zhang2019learning,cao2019end,xue2019building,liu2019cross,tu2019end}, however mostly for the purposes of code-mixing and cross-lingual synthesis. Language information was typically represented either with a language embedding~\cite{zhang2019learning,liu2019cross} or with a separate encoder for each language~\cite{nachmani2019unsupervised}, while~\cite{cao2019end} applied both approaches to code-mixing and accent conversion. With regards to multilingual modeling,~\cite{zhang2019learning} showed that multilingual models can attain a naturalness and speaker similarity that is comparable to that of a single speaker model for high-resource target languages, while research from~\cite{tu2019end} obtained promising results with a crosslingual transfer learning approach.

While research into S2S multilingual modeling is clearly vibrant, there appears to exist little systematic research into how S2S multilingual models could be used to increase speech naturalness for low-resource languages. To fill this void, this paper investigated to what extent results that are found in S2S monolingual multi-speaker modeling are transferable to multilingual multi-speaker modeling, and if it is possible to attain higher naturalness on low-resource languages with multilingual models than with single speaker models. Because multilingual modeling can benefit from the inclusion of large amounts of non-target language data, we also experimented with several data addition strategies and evaluated to what extent these strategies are effective to improve naturalness for low-resource languages. As this research is primarily addressing the viability of different approaches with regards to low-resource languages, our focus is not so much on maximizing naturalness but rather on gaining a better understanding of how different strategies work and would potentially scale up using larger amounts of data.

The rest of this paper is organized as follows.
In Section~\ref{sec:system}, we  describe the architecture used to conduct our experiments. In Section~\ref{sec:exp}, we describe the experimental design and give details about training and evaluation. In Section~\ref{sec:results}, we provide the experimental results. Finally, in Section~\ref{sec:conclusion}, we discuss conclusions and directions for future research.

\section{System architecture}\label{sec:system}

\subsection{S2S Acoustic model}

\begin{figure}[ht]
    \centering
    \includegraphics[width=\linewidth]{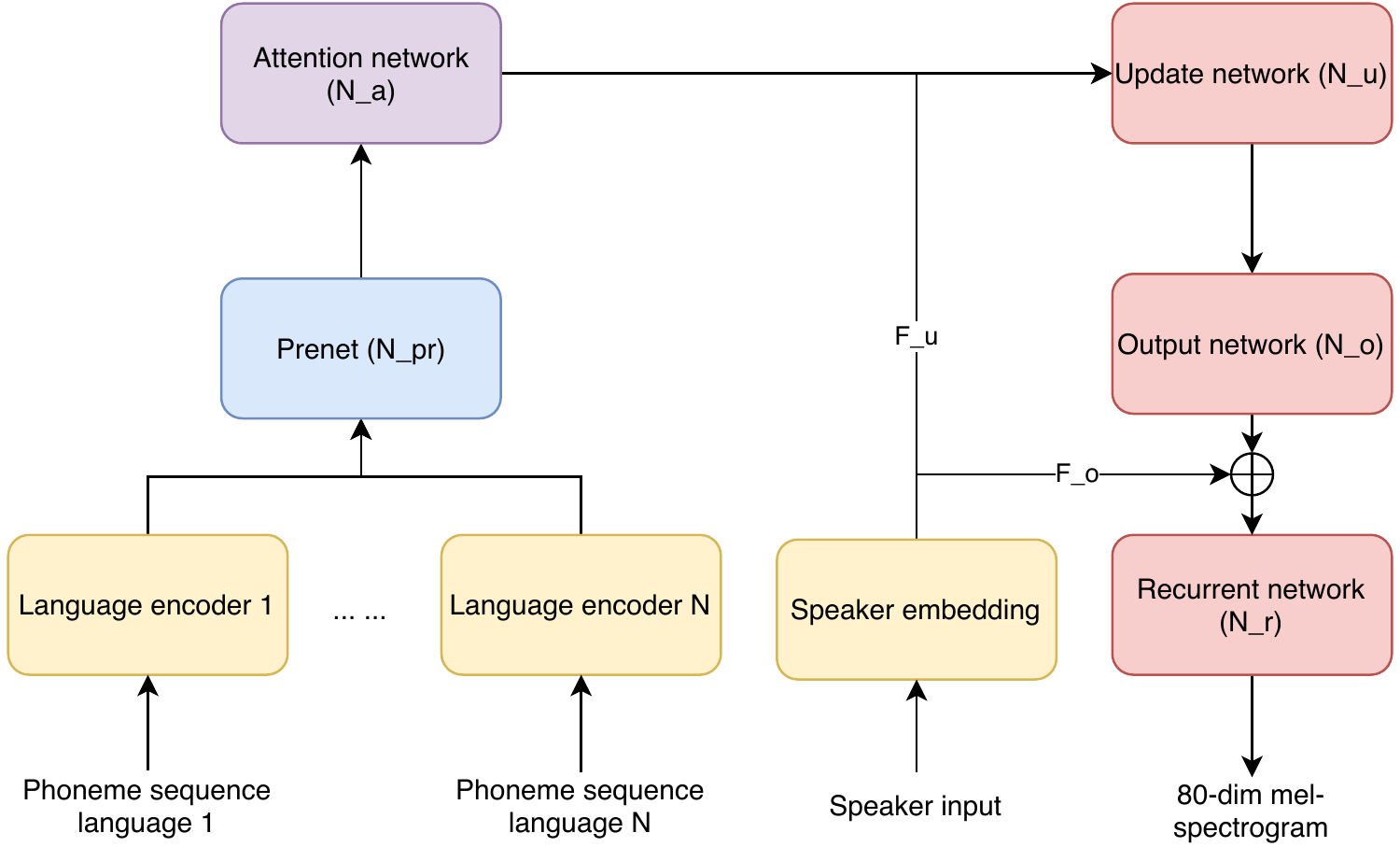}
    \caption{Overview of the acoustic model architecture used in this paper}
    \label{fig:architecture}
\end{figure}

The architecture that is used in this paper for acoustic modeling is based on VoiceLoop~\cite{taigman2017voiceloop}. This architecture is appealing for several reasons: the architecture is relatively small which makes it more suitable to train with smaller amounts of data, the model takes relatively little time to train, and it is capable of disentangling speaker information well for seen speakers~\cite{nachmani2018fitting}. To make the architecture suitable for multilingual modeling and increase its naturalness and robustness, we made several changes to the architecture. First, we incorporated a separate encoder for each language to disentangle language information, similar to~\cite{nachmani2019unsupervised}. We empirically found that representing language information this way was more effective than using a language embedding. This language encoder is used to convert phonemes from a language-dependent phone set into 256-dimensional embeddings. Second, we added a 3-layer convolutional prenet $N_{pr}$ in the style of~\cite{shen2018natural} to better model phonetic context. Third, we added a two-layer LSTM recurrency $N_r$ with 512 nodes to the decoder to better retain long-term information. The model was trained to produce 80-dimensional mel-spectrogram features in a way similar to ~\cite{shen2018natural}. The resulting architecture is visualized in Figure~\ref{fig:architecture}.

\subsubsection{Class weighted loss}
Training a multilingual model on a mixture of high-resource and low-resource languages can lead to class imbalances between languages, which can negatively affect naturalness for minority classes. Although it is possible to address this issue through over- and undersampling as explored in~\cite{luong2019training}, we instead decided to change the weighting of classes through our loss function, following~\cite{alumae2016improved}. The purpose of the reweighting is to increase importance of minority class samples, while reducing the impact of majority classes. The advantage of this approach is that the re-weighting operation has a low computational cost, and is therefore more efficient than oversampling or ensemble-based methods. The class weights were computed as follows:

\begin{equation}
\alpha_i = \sqrt{\frac{c}{c_i \times N}}
\end{equation}

Where $\alpha_i$ denotes the class weight that is computed for class $i$, $N$ refers to the number of classes, $c$ is the total number of samples, and $c_i$ is the number of samples for class $i$. It was suggested by~\cite{alumae2016improved}
that the model might become less robust if the variation in the class weights becomes too large. To counter this effect, we applied a square root operation to the weights and found that this led to better naturalness compared to both the unbalanced and the balanced weights. The weights were then normalized to correct for the square root operation, where $j$ is the index that iterates over the number of classes:

\begin{equation}
n\alpha_i = \alpha_i \times \frac{c}{\sum_{j}^{N} c_j \times \alpha_j}
\end{equation}

\section{Experimental setup}\label{sec:exp}

In this paper, we aimed to answer the following research questions:
\begin{enumerate}
    \item To what extent does adding data from non-target language speakers increase the naturalness for various amounts of data from a low-resource language?
    \item How does replacing monolingual multi-speaker models with multilingual multi-speaker models affect speech naturalness?
    \item In what way can additional non-target language data best be added to improve the naturalness of low-resource target language speech?

\end{enumerate}
Two listening experiments were designed to answer these research questions.

\subsection{Experimental design}
The first experiment was designed to compare the naturalness of single speaker models with that of multilingual models for different amounts of data from the target speaker. For this purpose, we trained three single speaker models using 2000, 4000, and 8000 sentences (referred to as SING-2k, SING-4k, and SING-8k respectively) as target language data. We also trained three multilingual models with the same amount of data for the target language, and added an additional 16000 sentences from a foreign language speaker (referred to as MULT-2k+16k, MULT-4k+16k, MULT-8k+16k). We hypothesize that the multilingual multi-speaker models will perform better than the single speaker models when the data set of the target speaker is limited, as we expect that the addition of foreign language data will improve the robustness of the model. We also hypothesize that the effect will become smaller when more target language data is available. We used the data of an American English speaker for the target language, and the data of a Dutch speaker as auxiliary language data. Other language pairs were tried internally to ensure that findings were reproducible. However, for the purposes of the listening test, American English was chosen as the target language to make subjective evaluation more straightforward, while Dutch was chosen to informally evaluate potential adverse effects in the auxiliary language.

The second experiment was designed to compare monolingual multi-speaker models to multilingual multi-speaker models, with similar as well as larger amounts of non-target language data. To evaluate how the models would behave when given similar amounts of data, we created a monolingual model (MONO-2k+16k) with 2000 sentences of our target speaker and 16000 sentences from another American English speaker. This model was compared to the MULT-2k+16k model that was also used in the previous experiment. We hypothesize that because of the effort to separate languages with language encoders, the multilingual model should attain a naturalness close to or similar to the naturalness of the monolingual model. Although there is more overlap in terms of pronunciation and prosody for monolingual speakers than for multilingual speakers, we expect that its effect on the naturalness of the target speaker should be limited because the rest of the model is trained jointly.

Because multilingual modeling makes it more straightforward to include language data from non-target languages, we also used this experiment to analyze whether adding more foreign language data could improve naturalness, and in which way additional data can best be added. We designed three additional models to evaluate this question. The first model, MULT-2k+2x16k, was trained on the same data as the MULT-2k+16k model, but with an additional 16000 sentences from a second Dutch speaker. If naturalness increases as a result of this additional data, this could indicate that it is beneficial for the model to have data from multiple speakers in the data set, for example to better separate speaker specific prosody and pronunciation patterns. The second model, MULT-2k+16k+16k, was trained on the same data as the MULT-2k+16k model, but with an additional 16000 sentences from a third language, in this case French. If naturalness increases significantly as a result of this strategy, it could be an indication that the model benefits from the ability to distinguish between large amounts of data, for example to better handle differences in prosody or pronunciation. The third model, MULT-2k+16x2k, was again trained on 2000 sentences from the target speaker, and an additional 2000 sentences from 16 speakers of 14 languages (13 European languages as well as Arabic). If this approach increases naturalness significantly, this could be an indication that the model benefits from language variety or from a lack of class imbalances.

To train the models, we used a proprietary text-to-speech data set. The speech consisted of recordings from professional voice talents who were asked to read aloud texts in a studio environment. After recording, all speech was processed and down-sampled to 22 kHz. Foreign language recordings, for example English recordings for non-English languages, were excluded to ensure that the results of this experiment were not impacted by such sentences.

For both experiments, we used a MUltiple Stimuli with Hidden Reference and Anchor (MUSHRA) test to evaluate naturalness~\cite{recommendation20011534}. Speaker similarity was not subjectively evaluated, because we found that the speaker characteristics of the target speaker were not harmed by the addition of data from other speakers. For the test, we recruited 30 participants with a good command of English. For both experiments, we created three separate test sets, each containing 10 stimulus panels with audio from unseen sentences. A participant was assigned one out of three test sets for both of the experiments, hence every participant evaluated 20 panels. This way, the time to complete the test was reduced whilst ensuring that results were not significantly impacted by a particular sentence. Following the MUSHRA guidelines, we included a resynthesized sample on each stimulus panel, both as a reference and as a hidden anchor.

For the design of the listening tests, we used the publicly available WebAudioEvaluationTool~\cite{jillings2015web}. Both the panels as well as the samples within a panel were randomized. In addition, the initial value of each slider in a panel was randomized to nudge participants to use the whole spectrum from 0 (completely unnatural) to 100 (completely natural). Participants had to listen to every sample and change the value of every slider before being allowed to proceed to the next panel. The experiments were then analyzed with a Wilcoxon signed-rank test, where a Holm-Bonferroni correction~\cite{holm1979simple} was used to reduce the chance of Type 1 errors.

\subsection{Training procedure}
The training of all acoustic models was done in two stages. Each model was first pretrained on sentences of up to 800 frames ($\approx$ 9.3 seconds), and split into separate parts up to 200 frames similar to~\cite{taigman2017voiceloop} to aid learning. For the pretraining, Stochastic Gradient Descent was used, with a batch size of 32, a learning rate of 0.1, and momentum of 0.75. After pretraining, the model was finetuned using the ADAM optimizer, with a batch size of 64, a learning rate of 0.0001 and betas of 0.9 and 0.98. For the monolingual multi-speaker models, class weighting was applied to the loss function to correct for imbalances in the speaker distribution. For the multilingual multi-speaker models, class weighting was applied to counter both speaker and language imbalances. The input to the models consisted of phonemes from a separate phoneset per language which were then converted into integers, while on the output side the models were trained to produce unnormalized 80-dimensional mel-spectrogram features. The mel-spectrogram features were then decoded by a WaveGlow vocoder~\cite{prenger2019waveglow}, that was trained in universal fashion~\cite{valle2019mellotron} on a proprietary data set consisting of 5000 sentences each from 3 female and 2 male speakers.

\section{Results}\label{sec:results}

\subsection{Experiment 1: Single-speaker modeling vs multilingual modeling}
For the first experiment, 30 participants were invited to evaluate 10 stimulus panels with 7 audio samples per panel. Of the 300 resulting data points, we discarded 15 data points where one single sample was rated considerably higher than the resynthesized sample. If multiple samples were rated higher than the resynthesized sample, we did not consider them anomalies and did not remove them. The rationale behind this approach is that if just a single sample was rated higher, it was more likely to be an outlier, and would also have a larger impact in the Wilcoxon rank testing than if multiple samples were rated higher.

\begin{figure}[h]
    \centering
    \includegraphics[width=\linewidth]{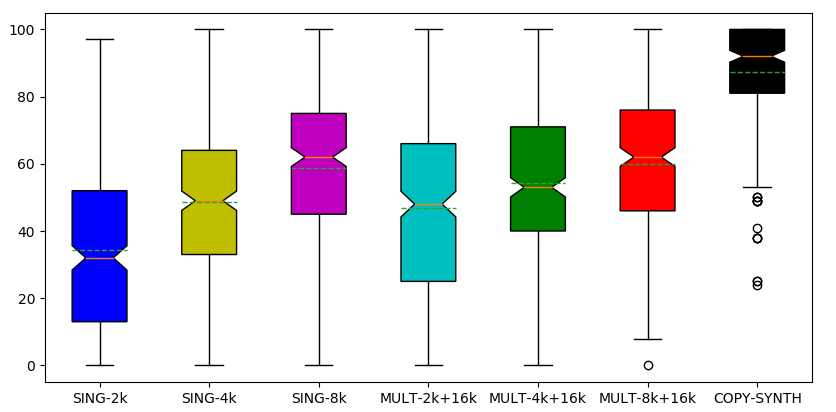}
    \caption{Boxplot showing the naturalness of single speaker models and multilingual models used in experiment 1. Red lines show median values, green lines show mean values}
    \label{fig:Experiment_1}
\end{figure}
The MUSHRA scores 
of the first experiment are displayed in Figure~\ref{fig:Experiment_1}. 
The results showed that the naturalness significantly increased when more target language data was available, both for single speaker and multilingual models. More interestingly, 
adding foreign language data to the target language data generally had a positive effect on the naturalness of the target speaker. When comparing the models with 2000 sentences of target language data, we found that the MULT-2k+16k model outperformed the SING-2k model significantly, and was on par with the SING-4k model ($p \approx 0.172$). Although the SING-2k model generally produced stable attention, the naturalness ratings for this model were negatively impacted by occasional mispronunciations that almost never occurred in the speech of other models. For the models that were trained on 4000 sentences from the low-resource language, the MULT-4k+16k model still produced significantly more natural speech than the SING-4k model. When comparing the models for which 8000 target language sentences were available, the 
difference in naturalness between the single speaker and the multilingual model was no longer significant 
($p \approx 0.506$).
All other system combinations were significantly different, and the resynthesized speech was rated significantly higher than speech from all other systems.

The results obtained for multilingual modeling followed similar patterns to the results in the monolingual multi-speaker settings in~\cite{latorre2019effect,luong2019training}. Similar to \cite{latorre2019effect}, the addition of non-target language data helped to improve the robustness and naturalness of the model when data quantities for the target language were limited, and similar to \cite{luong2019training}, the difference became insignificant when more target language data was available. The fact that the same effects could be replicated in a multilingual setting as in a monolingual setting suggests that the model does not suffer from being trained on different language inputs. We suspect that the effect is minimal because language information is well separated by the language encoders, thus limiting pronunciation overlap, while benefiting from shared training in the decoder.

\subsection{Experiment 2: Monolingual vs multilingual multi-speaker modeling}
Our second experiment was designed to better understand how various monolingual and multilingual model strategies may effect naturalness. We again asked 30 participants to evaluate ten different stimulus panels 
from one out of three test sets. Each panel consisted of a resynthesized sample as the reference and hidden anchor, and a sample from each of the five models. A similar procedure as in the first experiment was applied to remove anomalies, discarding 11 out of 300 data points.

\begin{table}[ht]
\caption{Subjective MUSHRA naturalness scores for systems in Experiment 2}
\centering
\begin{tabular}{lccc}
\toprule
\textbf{System identifier} & \textbf{Mean} & \textbf{Median} & \textbf{Average rank} \\ \midrule
MONO-2k+16k & 42.58 & 45 & 4.24 \\
MULT-2k+16k & 45.41 & 47 & 3.96 \\
MULT-2k+2x16k & 44.48 & 47 & 3.98 \\
MULT-2k+16k+16k & 45.51 & 48 & 3.91 \\
MULT-2k+16x2k & 47.24 & 50 & 3.72 \\
Resynthesis & 88.00 & 92 & 1.20 \\
\bottomrule
\end{tabular}
\label{tab:Experiment_2}
\end{table}

The results of the second experiment are displayed in Table~\ref{tab:Experiment_2}. When comparing the monolingual and the multilingual model that have similar amounts of data, we found that the multilingual MULT-2k+16k model performed on par with the monolingual MONO-2k+16k model ($p \approx 0.054$).
A significant difference between the monolingual model and multilingual models was found for some of the multilingual models with additional data, with a significant difference between the MONO-2k+16k and the MULT-2k+16x2k model ($p \approx 0.0003$), while the difference between the MONO-2k+16k and the MULT-2k+16k+16k model was marginally significant after Holm-Bonferroni correction ($p \approx 0.007$). Similar to the first experiment, the resynthesized speech was rated significantly better than the speech of all other systems. For the remaining system combinations, the differences were not significant. 

The results of this experiment showed that no significant difference in naturalness was found between the model with auxiliary target language data and the model with auxiliary non-target language data. We suspect that the difference is limited because the task of mel-spectrogram prediction is relatively language-indepedent. In fact, given that languages are separately modeled in the encoder, it might in some cases be beneficial to have auxiliary non-target language data instead of target language data because the architecture allows for better disentanglement of speaker-specific prosodic and pronunciation information.

When analyzing the multilingual models with additional data, we found 
that the naturalness of a multilingual model could even surpass that of a monolingual model, but that this was dependent on the sort of data added. While the MULT-2k+16x2k and the MULT-2k+16k+16k approach affected the naturalness of the target language positively, the MULT-2k+2x16k approach did not lead to a significant naturalness increase. These results thus suggest that when adding more data, a multilingual model benefits most from language variation and a reduction of class imbalances.

\section{Conclusions and Future Research}\label{sec:conclusion}

This paper aimed to investigate the effectiveness of multilingual modeling to improve speech naturalness of low-resource language neural speech synthesis. 
Our results showed 
that the addition of auxiliary non-target language data can positively impact the naturalness of low-resource language speech and can be a viable alternative to auxiliary target language data when such data is not readily available. We furthermore found that when more target language data was available, the inclusion of the auxiliary non-target language data did not negatively affect naturalness. Although 
we did not compare multilingual models with single speaker models for even larger amounts of target language data in this research, we expect that results from multilingual modeling will largely mimic the effects observed in studies of monolingual multi-speaker modeling~\cite{latorre2019effect}. Finally, we explored several strategies for including additional non-target language data. We showed that not all data addition strategies are equally effective, and reported that language diversity and minimizing class imbalances appear to be the most important variables to consider when adding data.

Based on our conclusions, we identify several directions for future research. First of all, the current research didn't consider the issue of language proximity on the effect of multilingual modeling. Although languages are modeled separately in the encoders, language proximity may positively affect naturalness. Additionally, this research evaluated low-resource language speech naturalness at a general level, while it may be more interesting to focus on the naturalness of language-specific characteristics such as language-specific phonemes or stress patterns. We furthermore note that the amount of auxiliary data used was relatively limited in our experiments. Further analysis could be done to find out whether our findings hold when scaled up with more data. Finally, we found that the MULT-2k+16x2k model was most effective to improve naturalness of target language speech, but this result does not clarify whether this effect can be attributed to the large variation in languages and speakers, or to the minimization of class imbalances. It would be interesting to disentangle these variables by comparing this model to a monolingual multi-speaker model with similar amounts of data per speaker.


\bibliographystyle{IEEEtran}
\bibliography{bibliography.bib}

\begin{thebibliography}{10}
\providecommand{\url}[1]{#1}
\csname url@samestyle\endcsname
\providecommand{\newblock}{\relax}
\providecommand{\bibinfo}[2]{#2}
\providecommand{\BIBentrySTDinterwordspacing}{\spaceskip=0pt\relax}
\providecommand{\BIBentryALTinterwordstretchfactor}{4}
\providecommand{\BIBentryALTinterwordspacing}{\spaceskip=\fontdimen2\font plus
\BIBentryALTinterwordstretchfactor\fontdimen3\font minus
  \fontdimen4\font\relax}
\providecommand{\BIBforeignlanguage}[2]{{%
\expandafter\ifx\csname l@#1\endcsname\relax
\typeout{** WARNING: IEEEtran.bst: No hyphenation pattern has been}%
\typeout{** loaded for the language `#1'. Using the pattern for}%
\typeout{** the default language instead.}%
\else
\language=\csname l@#1\endcsname
\fi
#2}}
\providecommand{\BIBdecl}{\relax}
\BIBdecl

\bibitem{shen2018natural}
J.~Shen, R.~Pang, R.~J. Weiss, M.~Schuster, N.~Jaitly, Z.~Yang, Z.~Chen,
  Y.~Zhang, Y.~Wang, R.~Skerrv-Ryan \emph{et~al.}, ``Natural tts synthesis by
  conditioning wavenet on mel spectrogram predictions,'' in \emph{2018 IEEE
  International Conference on Acoustics, Speech and Signal Processing
  (ICASSP)}.\hskip 1em plus 0.5em minus 0.4em\relax IEEE, 2018, pp. 4779--4783.

\bibitem{kons2019high}
Z.~Kons, S.~Shechtman, A.~Sorin, C.~Rabinovitz, and R.~Hoory, ``High quality,
  lightweight and adaptable tts using lpcnet,'' \emph{arXiv preprint
  arXiv:1905.00590}, 2019.

\bibitem{ren2019fastspeech}
Y.~Ren, Y.~Ruan, X.~Tan, T.~Qin, S.~Zhao, Z.~Zhao, and T.-Y. Liu, ``Fastspeech:
  Fast, robust and controllable text to speech,'' in \emph{Advances in Neural
  Information Processing Systems}, 2019, pp. 3165--3174.

\bibitem{jia2018transfer}
Y.~Jia, Y.~Zhang, R.~Weiss, Q.~Wang, J.~Shen, F.~Ren, P.~Nguyen, R.~Pang, I.~L.
  Moreno, Y.~Wu \emph{et~al.}, ``Transfer learning from speaker verification to
  multispeaker text-to-speech synthesis,'' in \emph{Advances in neural
  information processing systems}, 2018, pp. 4480--4490.

\bibitem{gibiansky2017deep}
A.~Gibiansky, S.~Arik, G.~Diamos, J.~Miller, K.~Peng, W.~Ping, J.~Raiman, and
  Y.~Zhou, ``Deep voice 2: Multi-speaker neural text-to-speech,'' in
  \emph{Advances in neural information processing systems}, 2017, pp.
  2962--2970.

\bibitem{deng2018modeling}
Y.~Deng, L.~He, and F.~Soong, ``Modeling multi-speaker latent space to improve
  neural tts: Quick enrolling new speaker and enhancing premium voice,''
  \emph{arXiv preprint arXiv:1812.05253}, 2018.

\bibitem{latorre2019effect}
J.~Latorre, J.~Lachowicz, J.~Lorenzo-Trueba, T.~Merritt, T.~Drugman,
  S.~Ronanki, and V.~Klimkov, ``Effect of data reduction on
  sequence-to-sequence neural tts,'' in \emph{ICASSP 2019-2019 IEEE
  International Conference on Acoustics, Speech and Signal Processing
  (ICASSP)}.\hskip 1em plus 0.5em minus 0.4em\relax IEEE, 2019, pp. 7075--7079.

\bibitem{luong2019training}
H.-T. Luong, X.~Wang, J.~Yamagishi, and N.~Nishizawa, ``Training multi-speaker
  neural text-to-speech systems using speaker-imbalanced speech corpora,''
  \emph{arXiv preprint arXiv:1904.00771}, 2019.

\bibitem{li2016multi}
B.~Li and H.~Zen, ``Multi-language multi-speaker acoustic modeling for lstm-rnn
  based statistical parametric speech synthesis,'' 2016.

\bibitem{gutkin2017uniform}
A.~Gutkin, ``Uniform multilingual multi-speaker acoustic model for statistical
  parametric speech synthesis of low-resourced languages,'' 2017.

\bibitem{nachmani2019unsupervised}
E.~Nachmani and L.~Wolf, ``Unsupervised polyglot text-to-speech,'' in
  \emph{ICASSP 2019-2019 IEEE International Conference on Acoustics, Speech and
  Signal Processing (ICASSP)}.\hskip 1em plus 0.5em minus 0.4em\relax IEEE,
  2019, pp. 7055--7059.

\bibitem{zhang2019learning}
Y.~Zhang, R.~J. Weiss, H.~Zen, Y.~Wu, Z.~Chen, R.~Skerry-Ryan, Y.~Jia,
  A.~Rosenberg, and B.~Ramabhadran, ``Learning to speak fluently in a foreign
  language: Multilingual speech synthesis and cross-language voice cloning,''
  \emph{arXiv preprint arXiv:1907.04448}, 2019.

\bibitem{cao2019end}
Y.~Cao, X.~Wu, S.~Liu, J.~Yu, X.~Li, Z.~Wu, X.~Liu, and H.~Meng, ``End-to-end
  code-switched tts with mix of monolingual recordings,'' in \emph{ICASSP
  2019-2019 IEEE International Conference on Acoustics, Speech and Signal
  Processing (ICASSP)}.\hskip 1em plus 0.5em minus 0.4em\relax IEEE, 2019, pp.
  6935--6939.

\bibitem{xue2019building}
L.~Xue, W.~Song, G.~Xu, L.~Xie, and Z.~Wu, ``Building a mixed-lingual neural
  tts system with only monolingual data,'' \emph{arXiv preprint
  arXiv:1904.06063}, 2019.

\bibitem{liu2019cross}
Z.~Liu and B.~Mak, ``Cross-lingual multi-speaker text-to-speech synthesis for
  voice cloning without using parallel corpus for unseen speakers,''
  \emph{arXiv preprint arXiv:1911.11601}, 2019.

\bibitem{tu2019end}
T.~Tu, Y.-J. Chen, C.-c. Yeh, and H.-y. Lee, ``End-to-end text-to-speech for
  low-resource languages by cross-lingual transfer learning,'' \emph{arXiv
  preprint arXiv:1904.06508}, 2019.

\bibitem{taigman2017voiceloop}
Y.~Taigman, L.~Wolf, A.~Polyak, and E.~Nachmani, ``Voiceloop: Voice fitting and
  synthesis via a phonological loop,'' \emph{arXiv preprint arXiv:1707.06588},
  2017.

\bibitem{nachmani2018fitting}
E.~Nachmani, A.~Polyak, Y.~Taigman, and L.~Wolf, ``Fitting new speakers based
  on a short untranscribed sample,'' \emph{arXiv preprint arXiv:1802.06984},
  2018.

\bibitem{alumae2016improved}
T.~Alum{\"a}e, S.~Tsakalidis, and R.~M. Schwartz, ``Improved multilingual
  training of stacked neural network acoustic models for low resource
  languages.'' in \emph{Interspeech}, 2016, pp. 3883--3887.

\bibitem{recommendation20011534}
I.~Recommendation, ``1534-1,“method for the subjective assessment of
  intermediate sound quality (mushra)”,'' \emph{International
  Telecommunications Union, Geneva, Switzerland}, 2001.

\bibitem{jillings2015web}
N.~Jillings, B.~Man, D.~Moffat, J.~D. Reiss \emph{et~al.}, ``Web audio
  evaluation tool: A browser-based listening test environment,'' 2015.

\bibitem{holm1979simple}
S.~Holm, ``A simple sequentially rejective multiple test procedure,''
  \emph{Scandinavian journal of statistics}, pp. 65--70, 1979.

\bibitem{prenger2019waveglow}
R.~Prenger, R.~Valle, and B.~Catanzaro, ``Waveglow: A flow-based generative
  network for speech synthesis,'' in \emph{ICASSP 2019-2019 IEEE International
  Conference on Acoustics, Speech and Signal Processing (ICASSP)}.\hskip 1em
  plus 0.5em minus 0.4em\relax IEEE, 2019, pp. 3617--3621.

\bibitem{valle2019mellotron}
R.~Valle, J.~Li, R.~Prenger, and B.~Catanzaro, ``Mellotron: Multispeaker
  expressive voice synthesis by conditioning on rhythm, pitch and global style
  tokens,'' \emph{arXiv preprint arXiv:1910.11997}, 2019.

\end{thebibliography}

\end{document}